\documentclass[pre,a4paper,showpacs,twocolumn,twoside]{revtex4}
\usepackage{times}
\usepackage{amsmath}
\usepackage{bm}
\def\onehalf{{\textstyle\frac{1}{2}}}
\def\quarter{{\textstyle\frac{1}{4}}}
\def\twothird{{\textstyle\frac{2}{3}}}
\def\vec#1{\bm{#1}}
\begin{document}
\title{Noether's theorem and Lie symmetries
for time-dependent Hamilton-Lagrange systems}
\author{J\"urgen Struckmeier}\email[Email address: ]{j.struckmeier@gsi.de}
\author{Claus Riedel}
\affiliation{Gesellschaft f\"ur Schwerionenforschung (GSI),
Planckstrasse~1, 64291~Darmstadt, Germany}
\received{8 May 2002; published 12 December 2002}
\begin{abstract}
Noether and Lie symmetry analyses based on point transformations
that depend on time and spatial coordinates will be reviewed for
a general class of time-dependent Hamiltonian systems.
The resulting symmetries are expressed in the form of generators
whose time-dependent coefficients follow as solutions of sets
of ordinary differential (``auxiliary'') equations.
The interrelation between the Noether and Lie sets of auxiliary
equations will be elucidated.
The auxiliary equations of the Noether approach will be shown
to admit invariants for a much broader class of potentials,
compared to earlier studies.
As an example, we work out the Noether and Lie symmetries
for the time-dependent Kepler system.
The Runge-Lenz vector of the time-independent Kepler system
will be shown to emerge as a Noether invariant if we
adequately interpret the pertaining auxiliary equation.
Furthermore, additional nonlocal invariants and symmetries
of the Kepler system will be isolated by identifying further
solutions of the auxiliary equations that depend on the
explicitly known solution path of the equations of motion.
Showing that the invariants remain unchanged under the action of
different symmetry operators, we demonstrate that a unique correlation
between a symmetry transformation and an invariant does not exist.
\end{abstract}
\maketitle
\section{\label{sec:intro}Introduction}
Analytical approaches aiming to analyze the particular properties
of a given dynamical system may successfully take advantage of the
formalism of infinitesimal symmetry transformations that have been
worked out by Lie~\cite{lie} and Noether~\cite{noether18}.
In this paper, we will review these approaches.
Specifically, both Noether and Lie symmetries will be
worked out on the basis of point transformations with
variations depending on time and spatial coordinates for a
general class of explicitly time-dependent Hamiltonian systems.
This parallel treatment will enable us to compare these symmetry
analyses, and to unveil both their close relationship and their
differences.
We will furthermore contribute to the ongoing discussion on how these
symmetries are related to the invariants of a given dynamical system.

The results of the symmetry analyses are obtained in the form
of generators of symmetry transformations.
The particular form of these generators is constituted by
time-dependent coefficients that are given as solutions of
ordinary differential (``auxiliary'') equations.
In order to obtain the full ``spectrum'' of these solutions,
the auxiliary equations {\em and\/} the system's equations
of motion must be conceived as a coupled
set~\cite{struck-rie00,struck-rie01,struck-rie02}.
The particular solutions of the auxiliary equations that
decouple from the solutions of the equations of motion
can then be seen to yield the generators of the
``fundamental'' system symmetries.

As an example, we work out the Noether and Lie symmetry
analyses for the time-dependent Kepler system.
The specific auxiliary equations are directly obtained
from the general formulation derived beforehand.
All known invariants and Lie symmetries will be shown
to emerge from the solutions of these auxiliary equations.
It is shown in particular that the Runge-Lenz vector of the
time-independent Kepler system is obtained as a classical
Noether invariant --- hence an invariant that arises from a
point transformation that depends on the vector of
spatial coordinates $\vec{q}$ and time $t$ --- if we
interpret the pertaining auxiliary equation appropriately.
Furthermore, not previously reported nonlocal Noether invariants
and Lie symmetries of the Kepler system will be isolated working
out additional solutions of the respective auxiliary equations.

The variation of the Noether invariants will be shown to vanish
under different Lie and Noether symmetry transformations.
We thereby demonstrate that a unique correlation between a
symmetry and a related invariant does not exist.

We start our analysis with a review of infinitesimal
point transformation and their generators in space-time.
This will be particularly helpful to clarify our notation and
to render our paper as self-contained as possible.
In this context, the brief presentation of Noether's
theorem will largely facilitate the understanding of
the Noether symmetry analysis of the general Hamiltonian
system that is governed by the potential $V(\vec{q},t)$,
as well as the subsequent Lie symmetry analysis.
\section{Infinitesimal point transformations}
Given a classical $n$-degree-of-freedom dynamical system
of particles, an infinitesimal point transformation denotes a
transformation that maps ``points'' in configuration space
and time into infinitesimal neighboring ``points'':
$(\vec{q},t)\mapsto(\vec{q}^{\,\prime}\!\!,t^{\prime})$,
the primes indicating the transformed quantities.
Formally, such a point transformation in the
$(\vec{q},t)$ space-time may be defined
in terms of an infinitesimal parameter $\varepsilon$ by
\begin{subequations}\label{infinitra}
\begin{alignat}{4}
t^{\prime}\,&=t&+&\,\delta t\,,&\qquad
&\delta t\,\,&=&\,\,\varepsilon\xi(\vec{q},t)\,,\label{infini-a}\\
q_{i}^{\prime}&=q_{i}\,&+&\,\delta q_{i}\,,&\qquad
&\delta q_{i}\,\,&=&\,\,\varepsilon\eta_{i}(\vec{q},t)\,.
\label{infini-b}
\end{alignat}
\end{subequations}
In order to derive the transformation rules for
$\dot{q}_{i}$ and $\ddot{q}_{i}$ for the infinitesimal
point transformation defined by Eqs.~(\ref{infini-a}) and
(\ref{infini-b}), we must be aware that the coordinates
$q_{i}$ and the time $t$ are transformed {\em simultaneously}.
The quantity $\delta\dot{q}_{i}$ follows from the
consideration that $\dot{q}_{i}^{\prime}$ is given by the
derivative of the transformed coordinate $q_{i}^{\prime}$
with respect to the transformed time $t^{\prime}$.
According to the transformation rules (\ref{infini-a})
and (\ref{infini-b}), we thus find~\cite{stephani}
\begin{eqnarray*}
\dot{q}_{i}^{\,\prime}=\frac{dq_{i}^{\prime}}{dt^{\prime}}&=&
\frac{dq_{i}+\varepsilon d\eta_{i}}{dt+\varepsilon d\xi}=
\frac{\dot{q}_{i}+\varepsilon\dot{\eta}_{i}}{1+\varepsilon\dot{\xi}}\\
&=&\dot{q}_{i}+\varepsilon\left(\dot{\eta}_{i}-\dot{\xi}\dot{q}_{i}
\right)+\boldsymbol{O}(\varepsilon^{2})\,,
\end{eqnarray*}
which means that the first-order variation
$\delta\dot{q}_{i}$ is given by
\begin{equation*}
\delta\dot{q}_{i}=\varepsilon\left[\dot{\eta}_{i}(\vec{q},t)-
\dot{\xi}(\vec{q},t)\,\dot{q}_{i}\right]\,.
\tag{\ref{infinitra}c}
\end{equation*}
The infinitesimal point transformation (\ref{infini-a})
and (\ref{infini-b}) thus uniquely determines the
mapping of the $\dot{q}_{i}^{\,\prime}$.
Similarly, we find the transformation rule for
the $\ddot{q}_{i}^{\,\prime}$ from
\begin{eqnarray*}
\ddot{q}_{i}^{\,\prime}=\frac{d\dot{q}_{i}^{\,\prime}}{dt^{\prime}}&=&
\frac{\ddot{q}_{i}+\varepsilon\left(\ddot{\eta}_{i}-\dot{\xi}
\ddot{q}_{i}-\ddot{\xi}\dot{q}_{i}\right)}{1+\varepsilon\dot{\xi}}\\
&=&\ddot{q}_{i}+\varepsilon\left(\ddot{\eta}_{i}-2\dot{\xi}
\ddot{q}_{i}-\ddot{\xi}\dot{q}_{i}\right)+\boldsymbol{O}(\varepsilon^{2})\,,
\end{eqnarray*}
which yields the variation $\delta\ddot{q}_{i}$ to first
order in $\varepsilon$,
\begin{equation*}
\delta\ddot{q}_{i}=\varepsilon\left(\ddot{\eta}_{i}-
2\dot{\xi}\,\ddot{q}_{i}-\ddot{\xi}\,\dot{q}_{i}\right)
\,.\tag{\ref{infinitra}d}
\end{equation*}
Given an arbitrary analytic function $u(\vec{q},t)$
of the $n$-di\-men\-sional vector of particle positions and time,
the function's variation $\delta u=u(\vec{q}^{\,\prime}\!\!,
t^{\prime})-u(\vec{q},t)$ that is induced by virtue of the
point transformation (\ref{infinitra}) is given by
\begin{displaymath}
\delta u=\frac{\partial u}{\partial t}\delta t+
\sum_{i=1}^{n}\frac{\partial u}{\partial q_{i}}\delta q_{i}=
\varepsilon\,\boldsymbol{U}\,u(\vec{q},t)\,,
\end{displaymath}
the operator $\boldsymbol{U}$ denoting the generator
of the infinitesimal point transformation (\ref{infinitra}),
\begin{equation}\label{generator}
\boldsymbol{U}=\xi(\vec{q},t)\frac{\partial}{\partial t}+\sum_{i=1}^{n}
\eta_{i}(\vec{q},t)\frac{\partial}{\partial q_{i}}\,.
\end{equation}
The variation $\delta v=v(\vec{q}^{\,\prime}\!\!,\dot{\vec{q}}\!\!
\phantom{q}^{\prime}\!\!,t^{\prime})-v(\vec{q},\dot{\vec{q}},t)$ of
an arbitrary analytic function $v(\vec{q},\dot{\vec{q}},t)$ follows as
\begin{displaymath}
\delta v=\frac{\partial v}{\partial t}\delta t+\sum_{i=1}^{n}
\left(\frac{\partial v}{\partial q_{i}}\delta q_{i}+
\frac{\partial v}{\partial\dot{q}_{i}}\delta\dot{q}_{i}\right)=
\varepsilon\,\boldsymbol{U}^{\prime}\,v(\vec{q},\dot{\vec{q}},t)\,,
\end{displaymath}
which means that the first ``extension'' $\boldsymbol{U}^{\prime}$
of the generator (\ref{generator}) is given by
\begin{equation}\label{1stext}
\boldsymbol{U}^{\prime}=\boldsymbol{U}+\sum_{i=1}^{n}
\eta_{i}^{\prime}\,\frac{\partial}{\partial\dot{q}_{i}}\,,\qquad
\eta_{i}^{\prime}=\dot{\eta}_{i}-\dot{\xi}\dot{q}_{i}\,.
\end{equation}
Finally, the variation $\delta w=w(\vec{q}^{\,\prime}\!\!,
\dot{\vec{q}}\!\!\phantom{q}^{\prime}\!\!,
\ddot{\vec{q}}\!\!\phantom{q}^{\prime}\!\!,t^{\prime})-
w(\vec{q},\dot{\vec{q}},\ddot{\vec{q}},t)$ of
an arbitrary analytic function
$w(\vec{q},\dot{\vec{q}},\ddot{\vec{q}},t)$ is obtained as
\begin{displaymath}
\delta w=\varepsilon\,\boldsymbol{U}^{\prime\prime}
\,w(\vec{q},\dot{\vec{q}},\ddot{\vec{q}},t)\,,
\end{displaymath}
with $\boldsymbol{U}^{\prime\prime}$ the second
``extension'' of the generator (\ref{generator}),
\begin{equation}\label{2ndext}
\boldsymbol{U}^{\prime\prime}=\boldsymbol{U}^{\prime}+\sum_{i=1}^{n}
\eta_{i}^{\prime\prime}\,\frac{\partial}{\partial\ddot{q}_{i}}\,,\quad
\eta_{i}^{\prime\prime}=\ddot{\eta}_{i}-2\dot{\xi}\ddot{q}_{i}-
\ddot{\xi}\dot{q}_{i}=\frac{d}{dt}\eta_{i}^{\prime}-\dot{\xi}\ddot{q}_{i}\,.
\end{equation}
We will make use of the second extension $\boldsymbol{U}^{\prime\prime}$
of the generator $\boldsymbol{U}$ in Sec.~\ref{sec:lie} for
a Lie symmetry analysis of a general time-dependent Lagrangian system.
Beforehand, the first extension $\boldsymbol{U}^{\prime}$ will be
needed in our review of Noether's theorem to be presented in the
following section.
\section{\label{sec:noether}Review of Noether's theorem}
Noether's theorem~\cite{noether18,hill51,lutzky78a} relates
the conserved quantities of an $n$-degree-of-freedom
Lagrangian system $L(\vec{q},\dot{\vec{q}},t)$ to
infinitesimal point transformations (\ref{infinitra})
that leave the Lagrange action $L\,dt$ invariant.
We now work out this theorem in the special form
that emerges from the infinitesimal point transformation
(\ref{infinitra}).
Among the general set of point transformations defined by
Eq.~(\ref{infinitra}), we consider exactly those that
leave the action $Ldt$ for a given Lagrangian
$L(\vec{q},\dot{\vec{q}},t)$ invariant,
\begin{equation}\label{lagact}
L\big(\vec{q},\dot{\vec{q}},t\big)\,dt\stackrel{!}{=}
L^{\prime}\big(\vec{q}^{\,\prime}\!\!,\dot{\vec{q}}\!\!
\phantom{q}^{\prime}\!\!,t^{\prime}\big)\,dt^{\prime}\,.
\end{equation}
Note that we allow the Lagrangian itself to change its
functional form by virtue of the point transformation
in order to satisfy the condition (\ref{lagact}).
As the system's equations of motion follow directly from the
variation of the action integral by virtue of Hamilton's
principle \mbox{$\delta\int Ldt=0$}, the condition (\ref{lagact})
implies the requirement that the particular symmetry transformation
(\ref{infinitra}) must sustain the form of the equations of motion.
This means that the point transformation (\ref{infinitra})
maps the action integral into another representation of
the {\em same\/} action integral.
In other words, we do {\em not\/} transform a physical system
into a different one, but regard a given Lagrangian system
$L(\vec{q},\dot{\vec{q}},t)$ from an infinitesimally dislodged
``viewpoint'' in order to isolate its inherent symmetries.

The {\em functional\/} relation between $L^{\prime}$ and $L$ may
be expressed introducing a gauge function $f(\vec{q},t)$,
\begin{equation}\label{LpL}
L^{\prime}\big(\vec{q}^{\,\prime}\!\!,\dot{\vec{q}}\!\!
\phantom{q}^{\prime}\!\!,t^{\prime}\big)=L+\delta L+\cdots=
L\big(\vec{q}^{\,\prime}\!\!,\dot{\vec{q}}\!\!
\phantom{q}^{\prime}\!\!,t^{\prime}\big)-\varepsilon
\frac{df}{dt}+\boldsymbol{O}(\varepsilon^{2})\,.
\end{equation}
For the relation (\ref{LpL}) to hold in general, it is necessary
and sufficient~\cite{hill51} that $f(\vec{q},t)$ depend on $\vec{q}$
and $t$ only since, according to Eq.~(\ref{infinitra}c),
the transformation $\dot{\vec{q}}\mapsto\dot{\vec{q}}
\!\!\phantom{q}^{\prime}$ is uniquely determined by
$\vec{q}\mapsto\vec{q}^{\,\prime}$ and $t\mapsto t^{\prime}$.
Inserting Eq.~(\ref{LpL}) into the condition for
the invariant Lagrange action (\ref{lagact}), we get
to first order in $\varepsilon$
\begin{equation}\label{lagact1}
L\big(\vec{q}^{\,\prime}\!\!,\dot{\vec{q}}\!\!
\phantom{q}^{\prime}\!\!,t^{\prime}\big)\,dt^{\prime}=
L\big(\vec{q},\,\dot{\vec{q}},t\big)\,dt+
\varepsilon\frac{df(\vec{q},t)}{dt}\,dt\,.
\end{equation}
On the other hand, the connection between
$L\big(\vec{q}^{\,\prime}\!\!,\dot{\vec{q}}\!\!
\phantom{q}^{\prime}\!\!,t^{\prime}\big)$ and
$L\big(\vec{q},\,\dot{\vec{q}},t\big)$ is determined by the
``extended'' operator $\boldsymbol{U}^{\prime}$ of Eq.~(\ref{1stext}),
\begin{displaymath}
L\big(\vec{q}^{\,\prime}\!\!,\dot{\vec{q}}\!\!
\phantom{q}^{\prime}\!\!,t^{\prime}\big)=
L\big(\vec{q},\dot{\vec{q}},t\big)+\varepsilon\,
\boldsymbol{U}^{\prime}\,L\big(\vec{q},\dot{\vec{q}},t\big)\,.
\end{displaymath}
To first order in $\varepsilon$,
Eq.~(\ref{lagact1}) thus yields the auxiliary
equation for $f(\vec{q},t)$, replacing $dt^{\prime}$
according to $dt^{\prime}=(1+\varepsilon\dot{\xi})\,dt$,
\begin{equation}\label{lagact2}
\frac{df(\vec{q},t)}{dt}=\dot{\xi}L+\boldsymbol{U}^{\prime}\,L\,.
\end{equation}
With the operators $\boldsymbol{U}$ and $\boldsymbol{U}^{\prime}$,
given by Eqs.~(\ref{generator}) and (\ref{1stext}), respectively,
the explicit form of Eq.~(\ref{lagact2}) reads
\begin{equation}\label{dotf}
\frac{df(\vec{q},t)}{dt}=\dot{\xi}\,L+
\xi\,\frac{\partial L}{\partial t}+\sum_{i=1}^{n}
\left(\eta_{i}\frac{\partial L}{\partial q_{i}}+
\big(\dot{\eta}_{i}-\dot{q}_{i}\dot{\xi}\big)
\frac{\partial L}{\partial \dot{q}_{i}}\right)\,.
\end{equation}
We may conceive Eq.~(\ref{dotf}) as a condition for the yet
unspecified functions $\xi(\vec{q},t)$ and $\eta_{i}(\vec{q},t)$.
Only those point transformations (\ref{infinitra})
whose constituents $\xi$ and $\eta_{i}$ satisfy
Eq.~(\ref{dotf}) maintain the Lagrange action $Ldt$
for the given Lagrangian $L(\vec{q},\dot{\vec{q}},t)$.

The terms of Eq.~(\ref{dotf}) can directly be split into
a total time derivative and a sum containing the Euler-Lagrange
equations of motion,
\begin{equation}\label{principle2}\begin{split}
\frac{d}{dt}&\left[f(\vec{q},t)-\xi\,L+
\sum_{i=1}^{n}\left(\xi\dot{q}_{i}-\eta_{i}\right)
\frac{\partial L}{\partial\dot{q}_{i}}\right]\\
&+\sum_{i=1}^{n}\left(\xi\dot{q}_{i}-\eta_{i}\right)
\left(\frac{\partial L}{\partial q_{i}}-
\frac{d}{dt}\frac{\partial L}{\partial\dot{q}_{i}}\right)=0\,.
\end{split}\end{equation}
Along the system trajectory $\big(\vec{q}(t),\dot{\vec{q}}(t)\big)$
given by the solutions of the Euler-Lagrange equations
\begin{equation}\label{euler-lagrange}
\frac{\partial L}{\partial q_{i}}-\frac{d}{dt}\frac{\partial L}
{\partial\dot{q}_{i}}=0\,,\qquad i=1,\dotsc,n\,,
\end{equation}
the related terms in Eq.~(\ref{principle2}) vanish.
This means that the time integral $I$ of the remaining terms
\begin{equation}\label{noether-theorem}
I=\sum_{i=1}^{n}\left(\xi\dot{q}_{i}-\eta_{i}\right)
\frac{\partial L}{\partial \dot{q}_{i}}-\xi L+f(\vec{q},t)
\end{equation}
constitutes a conserved quantity, i.e., a constant of motion
for the Lagrange system $L(\vec{q},\dot{\vec{q}},t)$.
The invariant given by Eq.~(\ref{noether-theorem}) together with
the differential equation (\ref{dotf}) for $f(\vec{q},t)$
is commonly referred to as Noether's theorem.
Starting from the initial condition
$(\vec{q}(t_{0}),\dot{\vec{q}}(t_{0}))$,
the system's state $(\vec{q}(t),\dot{\vec{q}}(t))$
is uniquely determined by the equations
of motion (\ref{euler-lagrange}), which in turn follow from
Hamilton's principle $\delta\int Ldt=0$.
Writing the variation $\delta\int L^{\prime}dt^{\prime}=0$
of the infinitesimally transformed system in terms of the original
coordinates, we obtain {\em in addition\/} to the equations
of motion (\ref{euler-lagrange}) the quantity $I$ that is
conserved by virtue of the symmetry transformation (\ref{infinitra}).
Thus, the requirement $\delta\int L^{\prime}dt^{\prime}=0$
may be seen as a generalization of Hamilton's principle that yields
both the equations of motion {\em and\/} a phase-space
symmetry relation embodied in the invariant $I$.
In general, Eq.~(\ref{dotf}) for $f(\vec{q},t)$
depends on $\vec{q}(t)$, hence on the solutions of the
equations of motion (\ref{euler-lagrange}).

Equation~(\ref{principle2}) exposes that the Noether
invariant (\ref{noether-theorem}) emerges simultaneously with
the time evolution of the system trajectory as the solution
of the Euler-Lagrange equations (\ref{euler-lagrange}).
Alternatively, Noether's theorem can be interpreted as a coupled
set of differential equations with the time $t$ the common
independent variable.
This coupled set consists of both the Euler-Lagrange equations of
motion~(\ref{euler-lagrange}) and an additional conditional
equation for $f(\vec{q},t)$.
In this regard, it can be considered as a generalized
Ermakov~\cite{ermakov} system whose time-dependent solutions
form together the invariant of Eq.~(\ref{noether-theorem}).
\section{Noether's theorem in Hamiltonian description}\label{sec:noeham}
From the definition of the Legendre transformation
\begin{equation}\label{H-L}
L(\vec{q},\dot{\vec{q}},t)=\sum_{i=1}^{n}p_{i}\,\dot{q}_{i}-
H(\vec{q},\vec{p},t)
\end{equation}
that maps a given Lagrangian $L(\vec{q},\dot{\vec{q}},t)$ into
the corresponding Hamiltonian $H(\vec{q},\vec{p},t)$, one
finds for the derivatives of $L$
\begin{equation}\label{H-L-deri}
\frac{\partial L}{\partial \dot{q}_{i}}=p_{i}\,,\qquad
\frac{\partial L}{\partial q_{i}}=-\frac{\partial H}{\partial q_{i}}\,,\qquad
\frac{\partial L}{\partial t}=-\frac{\partial H}{\partial t}\,.
\end{equation}
Applying these transformation rules for the transition from
a Lagrangian description of a dynamical system to a Hamiltonian
description to the Noether invariant of Eq.~(\ref{noether-theorem}),
one immediately gets
\begin{displaymath}
I=\sum_{i}^{n}\left(\xi\dot{q}_{i}-\eta_{i}\right)p_{i}-
\xi\sum_{i=1}^{n}p_{i}\,\dot{q}_{i}+\xi\,H+f(\vec{q},t)\,,
\end{displaymath}
which simplifies to the Hamiltonian formulation of Noether's theorem
\begin{equation}\label{ansatz}
I=\xi H-\sum_{i=1}^{n}\eta_{i}p_{i}+f(\vec{q},t)\,.
\end{equation}
The conditional equation for $f(\vec{q},t)$, given by
Eq.~(\ref{dotf}), translates according to Eqs.~(\ref{H-L-deri})
\begin{equation}\label{invder}
\dot{f}(\vec{q},t)=-\dot{\xi}H-\xi\,\frac{\partial H}{\partial t}+
\sum_{i=1}^{n}\left(\dot{\eta}_{i}p_{i}-
\eta_{i}\frac{\partial H}{\partial q_{i}}\right).
\end{equation}
Along the system trajectory, the canonical equations apply,
\begin{displaymath}
\frac{\partial H}{\partial p_{i}}=\frac{dq_{i}}{dt}\,,\qquad
\frac{\partial H}{\partial q_{i}}=-\frac{dp_{i}}{dt}\,,\qquad
\frac{\partial H}{\partial t}=\frac{dH}{dt}\,.
\end{displaymath}
Then, the right-hand side of Eq.~(\ref{invder}) can be expressed
as a total time derivative, yielding
\begin{displaymath}
\frac{d}{dt}\left(\xi H-
\sum_{i=1}^{n}\eta_{i}p_{i}+f(\vec{q},t)\right)=\frac{dI}{dt}=0.
\end{displaymath}
In the Hamiltonian formulation, Eq.~(\ref{dotf}) thus reduces to the
trivial statement that the total time derivative of the Noether invariant
$I$ from Eq.~(\ref{ansatz}) must vanish.
For a Hamiltonian $H$ with at most quadratic momentum dependence,
the form (\ref{ansatz}) of the Noether invariant is compatible with
an Ansatz function consisting of quadratic and linear terms in the
canonical momentum that has been used earlier by Lewis and
Leach~\cite{leach-lewis82}.
We thereby observe that this approach to work out an invariant is
mathematically equivalent to a strategy based on Noether's theorem
for this class of Hamiltonian systems.
\section{\label{sec:td-pot}Hamiltonian system with a general
time-dependent potential}
\subsection{Noether symmetry analysis}
To illustrate a particular Noether symmetry analysis,
we consider the $n$-degree-of-freedom system of particles
moving in an explicitly time-dependent potential $V(\vec{q},t)$,
\begin{equation}\label{ham2}
H(\vec{p},\vec{q},t)=\sum_{i=1}^{n}\onehalf p_{i}^{2}+V(\vec{q},t)\,.
\end{equation}
The canonical equations following from Eq.~(\ref{ham2}) are
\begin{equation}\label{eqmo2}
\dot{q}_{i}=\frac{\partial H}{\partial p_{i}}=p_{i}\,,\qquad
\dot{p}_{i}=-\frac{\partial H}{\partial q_{i}}=-\frac{\partial V}
{\partial q_{i}}\,.
\end{equation}
In the following, we work out the particular invariant $I$ of
the Hamiltonian system (\ref{ham2}) that specializes the general
Noether invariant in the form of Eq.~(\ref{ansatz}).
We hereby define a point mapping that is consistent with the
Noether symmetry transformation~(\ref{infinitra}).
For the particular Hamiltonian (\ref{ham2}), the general
condition for $dI/dt=0$ of Eq.~(\ref{invder}) reads
\begin{displaymath}
\frac{d}{dt}\!\!\left[\xi(\vec{q},t)\!\left(\sum_{i=1}^{n}\!\onehalf
p_{i}^{2}\!+\!V(\vec{q},t)\!\right)\!-\!\!\sum_{i=1}^{n}\!
\eta_{i}(\vec{q},t)\,p_{i}\!+\!f(\vec{q},t)\right]\!\!=\!0.
\end{displaymath}
Inserting the canonical equations (\ref{eqmo2}), the emerging
equation can only be fulfilled globally for any particular
vector of canonical momenta $\vec{p}$ if the sets of cubic,
quadratic, and linear momentum terms vanish separately ---
and correspondingly the sum of the remaining terms that do
not depend on the $p_{i}$,
\begin{subequations}\label{der2}
\begin{align}
\sum_{i}\sum_{j}\onehalf p_{i}^{2}p_{j}
\frac{\partial\xi}{\partial q_{j}}=&\;0\,,\label{der2a}\\
\sum_{i}\sum_{j}p_{i}p_{j}\left(\onehalf\delta_{ij}
\frac{\partial\xi}{\partial t}-\frac{\partial\eta_{i}}
{\partial q_{j}}\right)=&\;0\,,\label{der2b}\\
\sum_{i}p_{i}\left(\frac{\partial f}{\partial q_{i}}-
\frac{\partial\eta_{i}}{\partial t}+V\frac{\partial\xi}
{\partial q_{i}}\right)=&\;0\,,\label{der2c}\\
\sum_{i}\eta_{i}\frac{\partial V}{\partial q_{i}}+
\frac{\partial\xi}{\partial t}V+\xi\frac{\partial V}
{\partial t}+\frac{\partial f}{\partial t}=&\;0\,.\label{der2d}
\end{align}
\end{subequations}
The notation $\delta_{ij}$ in Eq.~(\ref{der2b}) stands for
the Kronecker symbol.
Since only a single momentum term appears in the sum of Eq.~(\ref{der2a}),
we may immediately conclude that the associated coefficient vanishes,
\begin{displaymath}
\frac{\partial\xi(\vec{q},t)}{\partial q_{j}}=0\,,
\qquad j=1,\dotsc,n\,,
\end{displaymath}
hence that $\xi(\vec{q},t)\equiv\beta(t)$ must be a function of $t$ only.
The double sum in Eq.~(\ref{der2b}) vanishes globally
for any $\vec{p}$ if $\partial\eta_{i}/\partial q_{j}$
cancels the $\dot{\xi}$ term up to a constant element
$a_{ij}$ of an antisymmetric matrix $(a_{ij})$
\begin{displaymath}
\frac{\partial\eta_{i}(\vec{q},t)}{\partial q_{j}}=
\onehalf\delta_{ij}\,\dot{\beta}(t)+a_{ij}\,,
\quad a_{ij}=-a_{ji}\,.
\end{displaymath}
In general form, the function $\eta_{i}(\vec{q},t)$
is thus given by
\begin{equation}\label{etai}
\eta_{i}(\vec{q},t)=\onehalf\dot{\beta}(t)\,q_{i}+\psi_{i}(t)+
\sum_{j=1}^{n}a_{ij}q_{j}\,.
\end{equation}
Herein, the $\psi_{i}(t)$ denote arbitrary functions of time only.
The linear momentum terms of Eq.~(\ref{der2c}) now require that
\begin{displaymath}
\frac{\partial f}{\partial q_{i}}=
\frac{\partial\eta_{i}}{\partial t}\,.
\end{displaymath}
Inserting the partial time derivative of Eq.~(\ref{etai}), we find
\begin{displaymath}
\frac{\partial f(\vec{q},t)}{\partial q_{i}}=
\onehalf\ddot{\beta}(t)\,q_{i}+\dot{\psi_{i}}(t)\,.
\end{displaymath}
This partial differential equation, too, may be generally
integrated to yield
\begin{equation}\label{f0}
f(\vec{q},t)=\ddot{\beta}(t)\sum_{i=1}^{n}\quarter
q_{i}^{2}+\sum_{i=1}^{n}\dot{\psi_{i}}(t)\,q_{i}\,.
\end{equation}
Now that $\eta_{i}$ and $f$ are specified by Eqs.~(\ref{etai}) and
(\ref{f0}), respectively, the invariant $I$ of Eq.~(\ref{ansatz})
can be expressed in terms of the yet unknown constants $a_{ij}$
and functions of time $\beta(t)$ and $\psi_{i}(t)$,
\begin{align}\label{inv2}
I=&\;\beta(t)\,H-\dot{\beta}(t)\sum_{i=1}^{n}\onehalf
q_{i}\,p_{i}+\ddot{\beta}(t)\sum_{i=1}^{n}\quarter q_{i}^{2}\nonumber\\
+&\;\sum_{i=1}^{n}\sum_{j=1}^{n}a_{ij}q_{i}\,p_{j}+
\sum_{i=1}^{n}\left(\dot{\psi}_{i}\,q_{i}-\psi_{i}\,p_{i}\right)\,.
\end{align}
The functions $\beta(t)$ and $\psi_{i}(t)$ and the $a_{ij}$ are
determined from condition~(\ref{der2d}), induced by the
terms not depending on the canonical momenta $p_{i}$,
\begin{equation}\label{der2r}
\dot{\beta}V+\beta\frac{\partial V}{\partial t}+\frac{\partial f}
{\partial t}+\sum_{i=1}^{n}\eta_{i}\frac{\partial V}{\partial q_{i}}=0\,.
\end{equation}
To obtain the explicit form of Eq.~(\ref{der2r}), we must insert
Eq.~(\ref{etai}) and the partial time derivative of Eq.~(\ref{f0}).
For potentials $V(\vec{q},t)$ that are not linear in $\vec{q}$,
the $\psi_{i}(t)$ terms are the only ones that depend linearly
on the canonical coordinates.
Consequently, the sum of these terms must vanish separately.
This means that two distinct differential equations are obtained,
namely for those terms that do {\em not\/} depend on $\psi_{i}(t)$
and the remaining terms that depend on $\psi_{i}(t)$.
The first group of terms of Eq.~(\ref{der2r}) form the following
inhomogeneous linear differential equation for $\beta(t)$,
keeping in mind that $a_{ij}=-a_{ji}$:
\begin{equation}\begin{split}\label{dgl1}
\dddot{\beta}(t)\sum_{i=1}^{n}\quarter q_{i}^{2}+&\;\dot{\beta}(t)
\left[V\big(\vec{q}(t),t\big)+\sum_{i=1}^{n}\onehalf q_{i}\frac{\partial V}
{\partial q_{i}}\right]\\
+&\;\beta(t)\frac{\partial V}{\partial t}+\sum_{i=1}^{n}
\sum_{j=1}^{n}a_{ij}\,q_{j}\frac{\partial V}
{\partial q_{i}}=0\,.
\end{split}\end{equation}
With $V(\vec{q}(t),t)$ the potential of Eq.~(\ref{ham2}),
Eq.~(\ref{dgl1}) represents an {\em ordinary\/} third-order
differential equation along the solution path $\vec{q}(t)$ of
the canonical equations~(\ref{eqmo2}) with the time $t$
the independent variable.
The general solution of Eq.~(\ref{dgl1}) is given by the
linear combination of its homogeneous part, together with a
particular solution of the inhomogeneous equation.
According to the existence and uniqueness theorem for linear
ordinary differential equations, a unique solution
$\beta(t)$ of the initial value problem~(\ref{dgl1}) exists
as long as its coefficients $\vec{q}(t)$, $V(\vec{q}(t),t)$,
and its partial derivatives are continuous along the
independent variable $t$.
Otherwise, the function $\beta(t)$ may cease to exist at some
finite instant of time $t_{1}$, which means that the related
invariant exists within the limited time span $t_{0}\le t<t_{1}$ only.

With our understanding of the auxiliary equation~(\ref{dgl1})
as an ordinary differential equation along the {\em known\/}
trajectory $\vec{q}(t)$, we differ from earlier studies
of Lewis and Leach~\cite{leach-lewis82}.
These authors conceived the auxiliary equation
as a partial differential equation for potentials $V(\vec{q},t)$.
Only those potentials that constitute a general solution
of Eq.~(\ref{dgl1}) were depicted to admit an invariant $I$.
We observe here that the invariant $I$ of Eq.~(\ref{inv2})
exists as well for the far more general class of potentials
$V(\vec{q}(t),t)$ that admit a solution $\beta(t)$ of
Eq.~(\ref{dgl1}) along the trajectory $\vec{q}(t)$.

For the functions $\psi_{i}(t)$,
Eq.~(\ref{der2r}) yields the condition
\begin{equation}\label{dgl2}
\sum_{i=1}^{n}\left(\ddot{\psi}_{i}(t)\,q_{i}+\psi_{i}(t)\,
\frac{\partial V}{\partial q_{i}}\right)=0\,.
\end{equation}
With $\psi_{i}(t)$ satisfying Eq.~(\ref{dgl2}), the
$\psi_{i}$-dependent terms of Eq.~(\ref{inv2})
form the separate invariant
\begin{equation}\label{inv4}
I_{\psi}=\sum_{i=1}^{n}\left[\dot{\psi}_{i}(t)\,q_{i}-
\psi_{i}(t)\,p_{i}\right]\,.
\end{equation}
We recall that the functions $\psi_{i}(t)$ emerge in
Eq.~(\ref{etai}) as separate integration ``constants''
for each index $i=1,\dotsc,n$.
Consequently, the invariant (\ref{inv4}) and
the related auxiliary equation (\ref{dgl2}) can be
split into a set of $n$ equations, respectively,
\begin{align}
\ddot{\psi}_{i}(t)\,q_{i}-\psi_{i}(t)\,
\dot{p}_{i}=0\,,\quad &i=1,\dotsc,n\,,\label{dgl2p}\\
I_{\psi_{i}}=\dot{\psi}_{i}(t)\,q_{i}-
\psi_{i}(t)\,p_{i}\,,\quad &i=1,\dotsc,n\,,\label{inv4p}
\end{align}
which means that the invariant $I$ can be written as a sum of
invariants $I=I_{\beta}+\sum_{i}I_{\psi_{i}}$.
The $\psi_{i}(t)$-independent terms of Eq.~(\ref{inv2}) thus form the
invariant $I_{\beta}$, which reads, inserting the Hamiltonian (\ref{ham2}),
\begin{equation}\begin{split}\label{inv3}
I_{\beta}=&\;\beta(t)\left[\sum_{i=1}^{n}\onehalf p_{i}^{2}+
V\big(\vec{q}(t),t\big)\right]-\dot{\beta}(t)
\sum_{i=1}^{n}\onehalf q_{i}\,p_{i}\\
+&\;\ddot{\beta}(t)\sum_{i=1}^{n}\quarter q_{i}^{2}+
\sum_{i=1}^{n}\sum_{j=1}^{n}a_{ij}\,q_{i}\,p_{j}\,.
\end{split}\end{equation}
With $\xi(\vec{q},t)\equiv\beta(t)$ and $\eta_{i}(\vec{q},t)$
given by Eq.~(\ref{etai}), the generators for the symmetry
transformations and their first extensions yielding the
Noether invariants (\ref{inv4}) and (\ref{inv3}) for the
class of Hamiltonian systems (\ref{ham2}) are given by
\begin{align}
\boldsymbol{U}=&\;\beta(t)\frac{\partial}{\partial t}+
\sum_{i}\bigg[\onehalf\dot{\beta}\,q_{i}+\sum_{j}a_{ij}
q_{j}+\psi_{i}(t)\bigg]\frac{\partial}{\partial q_{i}}\,,
\label{oper1}\\
\boldsymbol{U}^{\prime}=&\;\boldsymbol{U}+\sum_{i}\bigg[
\onehalf\ddot{\beta}\,q_{i}-\onehalf\dot{\beta}\,\dot{q}_{i}+
\sum_{j}a_{ij}\dot{q}_{j}+\dot{\psi}_{i}(t)\bigg]
\frac{\partial}{\partial\dot{q}_{i}}\notag\,.
\end{align}
Making use of the auxiliary equations (\ref{dgl1})
and (\ref{dgl2}) for $\beta(t)$ and the $\psi_{i}(t)$,
respectively, we may directly prove that
$\boldsymbol{U}^{\prime}$ satisfies the Noether requirement
(\ref{lagact2}) to yield a total time derivative of a
function $f(\vec{q},t)$ for the general class of Lagrangian
systems (\ref{H-L}) with the Hamiltonian of Eq.~(\ref{ham2}),
\begin{align*}
\boldsymbol{U}^{\prime}_{\!\psi}L=&\;\frac{d}{dt}\sum_{i}
\dot{\psi}_{i}(t)\,q_{i}\,,\quad\beta(t),a_{ij}\equiv0\,,\\
\boldsymbol{U}^{\prime}_{\!\beta}L+\dot{\xi}L=&\;\frac{d}{dt}
\ddot{\beta}(t)\sum_{i}\quarter q_{i}^{2}\,,\quad\psi_{i}(t)\equiv0\,,
\end{align*}
in agreement with Eq.~(\ref{f0}).
In order to verify that the variation $\delta I_{\beta}$ of
the Noether invariant (\ref{inv3}) indeed vanishes, we may
straightforwardly show that
\begin{displaymath}
\boldsymbol{U}^{\prime}_{\!\beta}I_{\beta}=0\,\,
\Longleftrightarrow\,\,\beta(t)\,\,
\text{is a solution of Eq.~(\ref{dgl1})}\,.
\end{displaymath}
With respect to the $\psi_{i}$-dependent part of
$\boldsymbol{U}^{\prime}$ acting on the invariant $I_{\psi_{i}}$
of Eq.~(\ref{inv4p}), we find, similarly,
\begin{align}\label{u-psi}
\boldsymbol{U}^{\prime}_{\!\psi_{i}}I_{\psi_{i}}=&\;\left(
\psi_{i}(t)\frac{\partial}{\partial q_{i}}+
\dot{\psi}_{i}(t)\frac{\partial}{\partial\dot{q}_{i}}\right)
\left[\dot{\psi}_{i}(t)\,q_{i}-\psi_{i}(t)\,\dot{q}_{i}
\right]\notag\\
=&\;\psi_{i}(t)\,\dot{\psi}_{i}(t)-
\dot{\psi}_{i}(t)\,\psi_{i}(t)\equiv0\,.
\end{align}
Obviously, the expression
$\boldsymbol{U}^{\prime}_{\!\psi_{i}}I_{\psi_{i}}$
vanishes separately for each index $i$, as it should be for
the $n$ distinct invariants $I_{\psi_{i}}$.
\subsection{\label{sec:lie}Lie symmetry analysis}
Another approach to the treatment of the symmetries of
a dynamical system has been established by Lie~\cite{lie}.
The class of Lie symmetries is defined by those
point transformations (\ref{infinitra}) that leave the
equations of motion invariant
\begin{equation}\label{eqmo3}
\ddot{q}_{i}+\frac{\partial V(\vec{q},t)}{\partial q_{i}}=0\,,
\qquad i=1,\dotsc,n\,.
\end{equation}
This coupled set of $n$ second-order equations
corresponds to the set of $2n$ first-order canonical
equations (\ref{eqmo2}).
As the equation of motion (\ref{eqmo3}) is of second order,
the condition for a vanishing variation reads
\begin{equation}\label{eqmo3a}
\boldsymbol{U}^{\prime\prime}\left(\ddot{q}_{i}+\frac{\partial
V(\vec{q},t)}{\partial q_{i}}\right)=0\,,\qquad i=1,\dotsc,n\,,
\end{equation}
with $\boldsymbol{U}^{\prime\prime}$ the second
extension~(\ref{2ndext}) of the generator $\boldsymbol{U}$.
Physically, a symmetry mapping of Eq.~(\ref{eqmo3})
that is associated with a vanishing variation (\ref{eqmo3a})
means to transform the equation of motion into the
{\em same\/} equation of motion in the new coordinate system.
Again, we thereby do not map our given physical system into
a different one, but isolate the conditions to be imposed
on the point mapping (\ref{infinitra}) in order to sustain
the form of the equation of motion.
As the particular dynamical system described by Eq.~(\ref{eqmo3})
is given in explicit form and does not involve velocity terms,
Eq.~(\ref{eqmo3a}) simplifies to
\begin{displaymath}
\eta_{i}^{\prime\prime}+\boldsymbol{U}\left(
\frac{\partial V(\vec{q},t)}{\partial q_{i}}\right)=0\,,
\end{displaymath}
which reads with $\boldsymbol{U}$ and $\eta_{i}^{\prime\prime}$
given by Eqs.~(\ref{generator}) and (\ref{2ndext})
\begin{displaymath}
\ddot{\eta}_{i}-2\dot{\xi}\ddot{q}_{i}-\ddot{\xi}\dot{q}_{i}+
\xi\frac{\partial^{2}V}{\partial q_{i}\partial t}+\sum_{j=1}^{n}
\eta_{j}\frac{\partial^{2}V}{\partial q_{i}\partial q_{j}}=0\,.
\end{displaymath}
Similar to the Noether symmetry analysis worked out in
Sec.~\ref{sec:noether}, this condition can only be fulfilled
globally for any velocity vector $\dot{\vec{q}}$ if and only
if the sets of linear, quadratic, and cubic velocity terms
vanish separately.
This requirement leads to the following hierarchy of
partial differential equations that must be fulfilled
for the given potential $V(\vec{q},t)$ by the functions
$\xi(\vec{q},t)$ and $\eta_{i}(\vec{q},t)$ of the
generator (\ref{generator})
\begin{subequations}\label{hierarchy}
\begin{align}
\sum_{j}\sum_{k}\dot{q}_{i}\dot{q}_{j}\dot{q}_{k}
\frac{\partial^{2}\xi}{\partial q_{j}\partial q_{k}}=&\;0,
\label{hie-a}\\
\sum_{j}\bigg[2\dot{q}_{i}\dot{q}_{j}
\frac{\partial^{2}\xi}{\partial q_{j}\partial t}-
\sum_{k}\dot{q}_{j}\dot{q}_{k}\frac{\partial^{2}\eta_{i}}
{\partial q_{j}\partial q_{k}}\bigg]=&\;0,\label{hie-b}\\
\sum_{j}\left[2\dot{q}_{j}\frac{\partial^{2}\eta_{i}}
{\partial q_{j}\partial t}+
\left(2\dot{q}_{j}\frac{\partial V}{\partial q_{i}}+
\dot{q}_{i}\frac{\partial V}{\partial q_{j}}\right)
\frac{\partial\xi}{\partial q_{j}}\right]-
\dot{q}_{i}\frac{\partial^{2}\xi}{\partial t^{2}}=&\;0,\label{hie-c}\\
\sum_{j}\!\!\left[\eta_{j}\frac{\partial^{2}V}{\partial q_{i}
\partial q_{j}}\!-\!\frac{\partial V}{\partial q_{j}}\frac{\partial\eta_{i}}
{\partial q_{j}}\right]\!+\!\frac{\partial^{2}\eta_{i}}{\partial t^{2}}+
2\frac{\partial V}{\partial q_{i}}\frac{\partial\xi}{\partial t}+
\xi\frac{\partial^{2} V}{\partial q_{i}\partial t}=&\;0.\label{hie-d}
\end{align}
\end{subequations}
Regarding Eq.~(\ref{hie-a}), we infer that all
second-order derivatives of $\xi(\vec{q},t)$ with respect
to the coordinates $q_{i}$ must be zero.
This means that $\xi(\vec{q},t)$ has the general form
\begin{equation}\label{lie-xi}
\xi(\vec{q},t)=\sum_{j}\alpha_{j}(t)\,q_{j}+\beta_{L}(t)\,,
\end{equation}
the $\alpha_{j}(t)$ and $\beta_{L}(t)$ denoting yet unknown
functions of time only.
The derivatives of $\xi(\vec{q},t)$ that are contained in
Eq.~(\ref{hierarchy}) may now be expressed as
\begin{align*}
\frac{\partial\xi}{\partial q_{j}}=&\;\alpha_{j}(t)\,,&
\frac{\partial\xi}{\partial t}=&\;\sum_{j}
\dot{\alpha}_{j}(t)\,q_{j}+\dot{\beta}_{L}(t)\,,\\
\frac{\partial^{2}\xi}{\partial q_{j}\partial t}=&\;
\dot{\alpha}_{j}(t)\,,&
\frac{\partial^{2}\xi}{\partial t^{2}}=&\;\sum_{j}
\ddot{\alpha}_{j}(t)\,q_{j}+\ddot{\beta}_{L}(t)\,.
\end{align*}
Equation~(\ref{hie-b}) is therefore globally fulfilled if
\begin{equation}\label{lie-eta0}
\frac{\partial^{2}\eta_{i}}{\partial q_{j}\partial q_{k}}=
\dot{\alpha}_{j}\delta_{ik}+\dot{\alpha}_{k}\delta_{ij}\,,
\end{equation}
the $\delta_{ik}$ and $\delta_{ij}$ meaning Kronecker symbols.
The general form of $\eta_{i}(\vec{q},t)$ is obtained from a
formal integration of Eq.~(\ref{lie-eta0}), introducing yet
undetermined functions of time $\gamma_{ij}(t)$ and $\phi_{i}(t)$,
\begin{equation}\label{lie-eta}
\eta_{i}(\vec{q},t)=\sum_{j}\left[\dot{\alpha}_{j}(t)\,q_{i}\,q_{j}+
\gamma_{ij}(t)\,q_{j}\right]+\phi_{i}(t)\,.
\end{equation}
The derivatives of $\eta(\vec{q},t)$ following from
Eq.~(\ref{lie-eta}) are
\begin{align*}
\frac{\partial\eta_{i}}{\partial q_{j}}=&\;\dot{\alpha}_{j}\,q_{i}+
\gamma_{ij}+\delta_{ij}\sum_{k}\dot{\alpha}_{k}\,q_{k}\,,\\
\frac{\partial^{2}\eta_{i}}{\partial q_{j}\partial t}=&\;
\ddot{\alpha}_{j}\,q_{i}+\dot{\gamma}_{ij}+
\delta_{ij}\sum_{k}\ddot{\alpha}_{k}\,q_{k}\,,\\
\frac{\partial^{2}\eta_{i}}{\partial t^{2}}=&\;\sum_{j}\left[
\dddot{\alpha}_{j}\,q_{i}\,q_{j}+\ddot{\gamma}_{ij}\,q_{j}
\right]+\ddot{\phi}_{i}\,.
\end{align*}
The conditions the time functions $\alpha_{j}(t)$, $\beta_{L}(t)$,
$\gamma_{ij}(t)$, and $\phi_{i}(t)$ must obey in order to
yield a valid symmetry transformation (\ref{eqmo3a})
are obtained inserting $\xi(\vec{q},t)$ of Eq.~(\ref{lie-xi})
and $\eta_{i}(\vec{q},t)$ from Eq.~(\ref{lie-eta}) together
with their respective partial derivatives into Eqs.~(\ref{hie-c})
and (\ref{hie-d}).
Distinguishing between terms that depend on $\vec{q}$ and those
that do not, the expression following from Eq.~(\ref{hie-c})
can be split into two sums that must vanish separately,
\begin{subequations}
\begin{align}
\sum_{j}\dot{q}_{j}\left(2\dot{\gamma}_{ij}-
\ddot{\beta}_{L}\,\delta_{ij}\right)=&\;0\label{betgam-rel}\,,\\
\sum_{j}\left[\ddot{\alpha}_{j}\left(2\dot{q}_{j}q_{i}+
\dot{q}_{i}q_{j}\right)+\alpha_{j}\left(2\dot{q}_{j}
\frac{\partial V}{\partial q_{i}}+\dot{q}_{i}\frac{\partial V}
{\partial q_{j}}\right)\right]=&\;0\label{alpha1}\,.
\end{align}
\end{subequations}
Equation~(\ref{betgam-rel}) can only be fulfilled globally
if $2\dot{\gamma}_{ij}=\ddot{\beta}_{L}\delta_{ij}$ for all
indices $i$ and $j$.
This means after time integration
\begin{equation}\label{betgam-rel1}
\gamma_{ij}(t)=\onehalf\dot{\beta}_{L}(t)\,\delta_{ij}+b_{ij}\,,
\end{equation}
with $b_{ij}$ denoting the integration constants.
With this result, $\eta_{i}(\vec{q},t)$
of Eq.~(\ref{lie-eta}) may be rewritten as
\begin{equation}\label{lie-eta1}
\eta_{i}(\vec{q},t)=\sum_{j}\dot{\alpha}_{j}(t)\,q_{i}\,q_{j}+
\sum_{j}\left[\onehalf\dot{\beta}_{L}(t)\,\delta_{ij}+b_{ij}\right]
q_{j}+\phi_{i}(t)\,.
\end{equation}
The $\vec{q}$-dependent terms of Eq.~(\ref{hie-c})
account for Eq.~(\ref{alpha1}).
Due to the coupling of the degrees of freedom that is induced
by the potential $V(\vec{q},t)$, it represents a set of $n$
auxiliary equations for the $n$ functions of time
$\alpha_{1}(t),\dotsc,\alpha_{n}(t)$.
Apart from particular potentials $V(\vec{q},t)$, this set may be
solved only along the system path $\vec{q}(t),\dot{\vec{q}}(t)$ that
emerges as the solution of the $n$ equations of motion (\ref{eqmo3}).

Finally, the terms of the hierarchy (\ref{hierarchy}) that
do not depend on $\dot{\vec{q}}$ must satisfy Eq.~(\ref{hie-d}).
Replacing $\gamma_{ij}$ according to Eq.~(\ref{betgam-rel1}),
we get three independent differential equations for the time functions
$\beta_{L}(t)$, $\phi_{i}(t)$, and $\alpha_{i}(t)$,
\begin{subequations}\label{lie-all}
\begin{align}
\dddot{\beta}_{L}q_{i}+\dot{\beta}_{L}\left[3\frac{\partial V}
{\partial q_{i}}+\sum_{j}q_{j}\frac{\partial^{2}V}
{\partial q_{i}\partial q_{j}}\right]+2\beta_{L}
\frac{\partial^{2}V}{\partial q_{i}\partial t}&\notag\\
\mbox{}-2\sum_{j}\left[b_{ij}\frac{\partial V}{\partial q_{j}}-
\frac{\partial^{2}V}{\partial q_{i}\partial q_{j}}
\sum_{k}b_{jk}\,q_{k}\right]=&\;0\,,\label{lie-beta}\\
\ddot{\phi}_{i}(t)+\sum_{j}\phi_{j}(t)\frac{\partial^{2}V}
{\partial q_{i}\partial q_{j}}=&\;0\,,\label{lie-phi}\\
\sum_{j}\left[\dddot{\alpha}_{j}\,q_{i}q_{j}+\dot{\alpha}_{j}
\left(q_{j}\frac{\partial V}{\partial q_{i}}-
q_{i}\frac{\partial V}{\partial q_{j}}\right)
\vphantom{\sum_{k}}\right.\quad\qquad\qquad&\nonumber\\\left.\mbox{}
+q_{j}\frac{\partial^{2}V}{\partial q_{i}\partial q_{j}}
\sum_{k}\dot{\alpha}_{k}\,q_{k}+\alpha_{j}\,q_{j}
\frac{\partial^{2}V}{\partial q_{i}\partial t}\right]
=&\;0\,.\label{alpha2}
\end{align}
\end{subequations}
As the degrees of freedom are coupled by the potential, each
equation stands for a set of $n$ coupled equations, with the
index ranging from $i=1,\dotsc,n$.
In terms of the solutions of the set of differential
equations (\ref{alpha1}) and (\ref{lie-all}), the
generator $\boldsymbol{U}_{\!L}$ of the symmetry
transformation (\ref{eqmo3a}) is given by
\begin{equation}\begin{split}\label{lie-oper1}
\boldsymbol{U}_{\!L}=&\;\bigg[\beta_{L}(t)+\sum_{i}
\alpha_{i}(t)\,q_{i}\bigg]\frac{\partial}{\partial t}\\
\mbox{}+\sum_{i}&\;\bigg[\onehalf\dot{\beta}_{L}q_{i}+
\sum_{j}b_{ij}\,q_{j}+\phi_{i}(t)+q_{i}\sum_{j}\dot{\alpha}_{j}
\,q_{j}\bigg]\frac{\partial}{\partial q_{i}}\,.
\end{split}\end{equation}
Obviously, this operator formally agrees  for $\alpha_{i}\equiv0$
with the generator (\ref{oper1}) of the Noether symmetry
transformation treated in Sec.~\ref{sec:td-pot}.
Nevertheless, we must keep in mind that the coefficients
of the operators (\ref{oper1}) and (\ref{lie-oper1}) are
different in general as they follow from a different set of
auxiliary equations.
Their interrelation becomes transparent considering
that Eqs.~(\ref{lie-beta}) and (\ref{lie-phi})
are partial $q_{i}$ derivatives of the
respective equations (\ref{dgl1}) and (\ref{dgl2})
of the Noether symmetry analysis.
Thus, Eqs.~(\ref{lie-beta}) and (\ref{lie-phi})
can be formally written as the partial $q_{i}$-derivative
of the Noether condition (\ref{lagact2})
\begin{equation}\label{brief}
\frac{\partial}{\partial q_{i}}\left[\boldsymbol{U}_{\!L}^{\prime}
\,L+\dot{\beta}_{L}(t)\,L-\frac{df_{L}(\vec{q},t)}{dt}\right]=0\,,
\end{equation}
the operator $\boldsymbol{U}_{\!L}^{\prime}$ given by the first
extension of Eq.~(\ref{lie-oper1}), the Lagrangian $L$ by
Eqs.~(\ref{H-L}) and (\ref{ham2}), and the particular gauge
function $f_{L}$ for the actual system that corresponds
to Eq.~(\ref{f0}) given by
\begin{displaymath}
f_{L}(\vec{q},t)=\ddot{\beta}_{L}(t)\sum_{i=1}^{n}\quarter
q_{i}^{2}+\sum_{i=1}^{n}\dot{\phi_{i}}(t)\,q_{i}\,.
\end{displaymath}
Regarding the homogeneous equations for the $\alpha_{i}$,
we observe that Eqs.~(\ref{alpha1}) and (\ref{alpha2}) impose
a set of $2n$ conditions for the $n$ functions of time
$\alpha_{i}(t)$.
We conclude that --- apart from very specific potentials
$V(\vec{q},t)$ --- these conditions cannot be satisfied.
This means that in most cases Eqs.~(\ref{alpha1}) and
(\ref{alpha2}) admit the trivial solution
$\vec{\alpha}(t)\equiv0$ only, hence no
$\vec{\alpha}$-related Lie symmetries exist.
The one-dimensional time-dependent harmonic-oscillator
system is one exception.
It is easily shown that Eqs.~(\ref{alpha1}) and (\ref{alpha2})
are compatible for this particular system, leading to the
well-known additional Lie symmetries~\cite{wulf} that exist
in addition to the Noether symmetries.
In Sec.~\ref{sec:example}, we demonstrate that these
equations also admit a nontrivial solution for the
Kepler system --- yielding a yet unreported Lie symmetry
of this system.
Kepler's third law is shown to originate from a particular
solution of Eq.~(\ref{lie-beta}) for $\beta_{L}(t)$.
We will furthermore show that the familiar invariants given by
the energy conservation law, the conservation of the angular
momentum, and the Runge-Lenz vector are Noether symmetries.
Finally, two new Noether invariants for the Kepler system
are derived from the solutions of Eq.~(\ref{dgl1}) with
$\beta(t)\ne\text{const}$.
\section{\label{sec:example}Example: Kepler system}
\subsection{Equation of motion}
The classical Kepler system is a two-body problem with
the mutual interaction following an inverse square force law.
In the frame of the reference body, the Cartesian coordinates
$q_{1}, q_{2}$ of its counterpart may be described in the plane
of motion by
\begin{equation}\label{kepler}
\ddot{q}_{i}+\mu(t)\frac{q_{i}}
{\sqrt{{\big(q_{1}^{2}+q_{2}^{2}\big)}^{3}}}=0\,,\qquad i=1,2\,,
\end{equation}
with $\mu(t)=G\big[m_{1}(t)+m_{2}(t)\big]$ the time-dependent
gravitational coupling strength that is induced by time-dependent
masses $m_{1}(t)$ and $m_{2}(t)$ of the interacting bodies.
We may regard the equation of motion (\ref{kepler}) to
originate from the Hamiltonian
\begin{equation}\label{kepham}
H(\vec{q},\vec{p},t)=\onehalf p_{1}^{2}+
\onehalf p_{2}^{2}+V(\vec{q},t)
\end{equation}
containing the interaction potential
\begin{equation}\label{keppot}
V(\vec{q},t)=-\frac{\mu(t)}{\sqrt{q_{1}^{2}+q_{2}^{2}}}=
-\frac{\mu(t)}{r}\,.
\end{equation}
\subsection{\label{sec:noe-kep}Noether symmetry analysis}
The complete set of Noether invariants (\ref{inv4p}) and
(\ref{inv3}) together with its related generator (\ref{oper1})
is obtained by finding the complete set of
solutions of the differential equations (\ref{dgl1}) and
(\ref{dgl2p}) for the particular potential (\ref{keppot}).
\subsubsection{Solutions related to $\beta(t)$ and $a_{ij}$}
We start with the inhomogeneous part of Eq.~(\ref{dgl1})
originating from a nonvanishing antisymmetric matrix
$(a_{ij})$ that is contained in the general solution
of Eq.~(\ref{der2b}).
For our actual two-dimensional system, this matrix cannot
contain more than one independent element, viz.\
$a_{11}=a_{22}=0$, $a_{12}=-a_{21}$.
The double sum of Eq.~(\ref{dgl1}) thus reads, explicitly,
\begin{align*}
\sum_{i=1}^{2}\sum_{j=1}^{2}a_{ij}\,q_{j}
\frac{\partial V}{\partial q_{i}}=&\;a_{12}\,q_{2}
\frac{\mu q_{1}}{r^{3}}+a_{21}\,q_{1}\frac{\mu q_{2}}{r^{3}}\\
=&\;\frac{\mu}{r^{3}}a_{12}\left(q_{1}q_{2}-q_{2}q_{1}\right)
\equiv0
\end{align*}
for arbitrary constants $a_{12}$.
Therefore, the auxiliary equation (\ref{dgl1}) has the
nontrivial solution $a_{12}\ne 0$, independently of $\beta(t)$.
Defining $a_{12}=1$, we thus obtain the separate invariant
$I_{a}$ from Eq.~(\ref{inv3}),
\begin{equation}\label{dreh}
I_{a}= q_{1}\,p_{2}-q_{2}\,p_{1}\,.
\end{equation}
Obviously, this invariant represents Kepler's second law,
stating that the angular momentum is a conserved quantity.
The associated generator $\boldsymbol{U}_{a}$ of the
symmetry transformation is readily obtained from Eq.~(\ref{oper1})
for $a_{12}=-a_{21}=1$,
\begin{equation}\label{drehgen}
\boldsymbol{U}_{a}=q_{2}\frac{\partial}{\partial q_{1}}-
q_{1}\frac{\partial}{\partial q_{2}}\,.
\end{equation}
The homogeneous part of Eq.~(\ref{dgl1}) forms a
separate auxiliary equation for $\beta(t)$.
For our given potential of Eq.~(\ref{keppot}), we find
the third-order equation
\begin{equation}\label{betakep}
\dddot{\beta}(t)-\dot{\beta}(t)\frac{2\mu(t)}{r^{3}(t)}-
\beta(t)\frac{4\dot{\mu}(t)}{r^{3}(t)}=0\,.
\end{equation}
With the Hamiltonian (\ref{kepham}), and $\beta(t)$
a solution of Eq.~(\ref{betakep}), the associated
invariant $I_{\beta}$ is given by
\begin{equation}\label{energy}
I_{\beta}=\beta(t)\,H-\onehalf\dot{\beta}(t)\left(
q_{1}p_{1}+q_{2}p_{2}\right)+\quarter\ddot{\beta}(t)
\left(q_{1}^{2}+q_{2}^{2}\right)\,.
\end{equation}
The generators of the symmetry transformations pertaining to
the three linear independent solutions of Eq.~(\ref{betakep})
follow from Eq.~(\ref{oper1}) as
\begin{equation}\label{betagen}
\boldsymbol{U}_{\beta_{i}}=\beta_{i}(t)\frac{\partial}
{\partial t}+\onehalf\dot{\beta}_{i}(t)\left(q_{1}
\frac{\partial}{\partial q_{1}}+q_{2}\frac{\partial}
{\partial q_{2}}\right)\,.
\end{equation}
For the conventional case of a {\em constant\/} coupling strength
[\mbox{$\dot{\mu}(t)=0$}], the auxiliary equation~(\ref{betakep})
has the particular solution $\beta_{1}(t)=1$.
With this solution, the invariant (\ref{energy}) reduces to
\begin{displaymath}
I_{\beta_{1}=1}=H\,,
\end{displaymath}
which provides the familiar result that the instantaneous system
energy $H$ that is given by the Hamiltonian (\ref{kepham}) is a
conserved quantity if $H$ does not depend on time explicitly.
The generator of the corresponding symmetry transformation
then simplifies to
\begin{equation}\label{energen}
\boldsymbol{U}_{\beta_{1}=1}=\frac{\partial}{\partial t}\,.
\end{equation}
As for all time-independent systems, two nonconstant
solutions $\beta_{2,3}(t)$ of Eq.~(\ref{betakep}) always exist.
For the Kepler system with $\dot{\mu}=0$, these solutions
can be expressed as
\begin{displaymath}
\beta_{2,3}(t)=\int_{t_{0}}^{t}
\frac{u\big(\theta(\tau)\big)\,d\tau}{1+\varepsilon\cos{\theta}(\tau)}\,,
\end{displaymath}
where the function $u(\theta)$ is one of the two
solutions of the differential equation,
\begin{displaymath}
\frac{d^{2}u}{d\theta^{2}}+\left(1-\frac{3}
{1+\varepsilon\cos{\theta}}\right)\,u(\theta)=0\,,
\end{displaymath}
and $\theta(t)$ the polar angle of the elliptical
trajectory with eccentricity $\varepsilon$.
These independent solutions $\beta_{2,3}(t)$ induce two
additional nonlocal invariants $I_{\beta_{2,3}}$ of the
form of Eq.~(\ref{energy}) which --- to the authors'
knowledge --- have not been previously reported.
\subsubsection{Solutions related to $\psi_{i}(t)$}
The $\psi_{i}$-related invariants (\ref{inv4p}) are obtained
from the solutions of the auxiliary equations (\ref{dgl2p}).
Inserting our given equation of motion (\ref{kepler}) into
Eq.~(\ref{dgl2p}), we find
\begin{equation}\label{psikep}
\ddot{\psi}_{i}(t)+\psi_{i}(t)\frac{\mu(t)}{r^{3}(t)}=0\,.
\end{equation}
With $\psi_{i}(t)$ and $\dot{\psi}_{i}(t)$ a solution
of the auxiliary equation (\ref{psikep}),
the associated Noether invariants (\ref{inv4p}) read
\begin{equation}\label{runge}
I_{\psi_{i}}=\dot{\psi}_{i}(t)\,q_{i}-
\psi_{i}(t)\,\dot{q}_{i}\,,\quad i=1,2\,.
\end{equation}
The two independent generators of the symmetry transformation
that result from the two linear independent solutions of
Eq.~(\ref{psikep}) are
\begin{equation}\label{rlgen}
\boldsymbol{U}_{\psi_{i}}=\psi_{i}(t)
\frac{\partial}{\partial q_{i}}\,,\quad i=1,2\,.
\end{equation}
It is again instructive to contemplate in particular
the time-independent case.
We may easily convince ourselves by direct insertion that
\begin{equation}\label{psisol}
\psi_{i}(t)=q_{1}(t)\,\dot{q}_{1}(t)+q_{2}(t)\,\dot{q}_{2}(t)
\end{equation}
is a solution of Eq.~(\ref{psikep}) provided that $\dot{\mu}(t)=0$.
Inserting Eq.~(\ref{psisol}) and its total time derivative
\begin{displaymath}
\dot{\psi}_{i}(t)=\dot{q}_{1}^{2}(t)+\dot{q}_{2}^{2}(t)-
\frac{\mu}{r(t)}
\end{displaymath}
into Eq.~(\ref{runge}), the invariants read, explicitly,
\begin{subequations}\label{runge-all}
\begin{align}
I_{\psi_{1}}=&\;q_{1}\dot{q}_{2}^{2}-q_{2}\dot{q}_{1}
\dot{q}_{2}-q_{1}\frac{\mu}{r}\,,\\
I_{\psi_{2}}=&\;q_{2}\dot{q}_{1}^{2}-q_{1}\dot{q}_{1}
\dot{q}_{2}-q_{2}\frac{\mu}{r}\,.
\end{align}
\end{subequations}
Obviously, the Noether invariants (\ref{runge-all}) represent
the two components of the Runge-Lenz vector.
This result contrasts with the usual perception of the
Runge-Lenz vector as a ``non-Noether invariant''~\cite{prince}.
Nevertheless, we must be very careful writing the Noether
invariants (\ref{runge-all}) in this form.
The requirement
\mbox{$\boldsymbol{U}^{\prime}_{\psi_{i}}I_{\psi_{i}}=0$}
of Eq.~(\ref{u-psi}) for the first extension of the generator
(\ref{rlgen}) acting on the invariant (\ref{runge}) is satisfied
if and only if the invariant is written in the form of
Eq.~(\ref{runge}) with $\psi_{i}(t)$ given by Eq.~(\ref{psisol}).
Only in this form is the right distinction between spatial and time
dependence made in Eq.~(\ref{runge}) --- with $\psi_{i}(t)$
written as a function of time only that is defined along the solution
path $(\vec{q}(t),\dot{\vec{q}}(t))$ of the equations of motion.
\subsection{Lie symmetry analysis}
Similar to the Noether analysis, we may systematically isolate
the complete set of Lie symmetries of the equation of
motion~(\ref{kepler}) by finding all solutions of the
auxiliary equations (\ref{alpha1}) and (\ref{lie-all})
for the coefficients $\vec{\alpha}(t)$, $\beta_{L}(t)$,
$\vec{\phi}(t)$, and the constant matrix $(b_{ij})$ that
constitute the Lie generator (\ref{lie-oper1}).
\subsubsection{Solutions related to $\alpha_{i}(t)$}
We start our Lie analysis with the time functions $\alpha_{i}(t)$,
given as the simultaneous solutions of Eqs.~(\ref{alpha1}) and
(\ref{alpha2}).
With Eq.~(\ref{keppot}) the potential of the Kepler system,
the condition (\ref{alpha1}) takes on the particular form
\begin{equation}\label{alphasol1}
\ddot{\alpha}_{i}(t)+\alpha_{i}(t)\frac{\mu}{r^{3}}=0\,,\quad i=1,2\,.
\end{equation}
The total time derivative of Eq.~(\ref{alphasol1}) inserted
into Eq.~(\ref{alpha2}) then provides the condition for
Eqs.~(\ref{alpha1}) and (\ref{alpha2}) to be simultaneously satisfied,
\begin{displaymath}
\big(\dot{\alpha}_{1}\,q_{1}+\dot{\alpha}_{2}\,q_{2}\big)
\big(q_{1}^{2}+q_{2}^{2}\big)=\big(\alpha_{1}q_{1}+
\alpha_{2}q_{2}\big)\big(q_{1}\dot{q}_{1}+q_{2}\dot{q}_{2}\big)\,.
\end{displaymath}
The obvious solution is to identify the time functions
$\alpha_{i}(t)$ with the time evolution of the coordinates $q_{i}(t)$,
\begin{displaymath}
\alpha_{i}(t)=c\,q_{i}(t)\,,\quad i=1,2\,.
\end{displaymath}
The related generator of this Lie symmetry reads
\begin{equation}\begin{split}\label{alphasol2}
\boldsymbol{U}_{L,\alpha}=\big[&\alpha_{1}(t)\,q_{1}+
\alpha_{2}(t)\,q_{2}\big]\,\frac{\partial}{\partial t}\\
\mbox{}+\big[&\dot{\alpha}_{1}(t)\,q_{1}+\dot{\alpha}_{2}(t)\,q_{2}\big]
\left(q_{1}\frac{\partial}{\partial q_{1}}+q_{2}
\frac{\partial}{\partial q_{2}}\right)\,.
\end{split}\end{equation}
We note that the identification of the time evolution
of $\alpha_{i}(t)$ with the time evolution of the
spatial coordinates $q_{i}(t)$ holds for {\em arbitrary\/}
time evolutions $\mu(t)$ of the coupling strength.
The question whether a physical interpretation of this
yet unreported Lie symmetry of the Kepler system exists
must be left unanswered at this point.
However, an interesting connection between the Lie symmetry
generator (\ref{alphasol2}) and the Noether invariants of
Eqs.~(\ref{dreh}) and (\ref{energy}) is revealed by letting
the first extension of the operator (\ref{alphasol2})
act on $I_{a}$ and $I_{\beta}$, respectively.
Provided that $\beta(t)$ is a solution of
Eq.~(\ref{betakep}), we find
\begin{displaymath}
\boldsymbol{U}^{\prime}_{L,\alpha}I_{a}=0\,,\quad
\boldsymbol{U}^{\prime}_{L,\alpha}I_{\beta}=0\,.
\end{displaymath}
Obviously, the Noether invariants $I_{a}$ and $I_{\beta}$
are also invariants with respect to the Lie symmetry that
is generated by Eq.~(\ref{alphasol2}).
This shows that a unique correlation of invariants and
symmetry operators is not possible.
\subsubsection{Solutions related to $\beta_{L}(t)$ and $b_{ij}$}
In the next step, we work out the set of solutions $\beta_{L}(t)$
of Eq.~(\ref{lie-beta}) for the potential~(\ref{keppot}).
In our particular case, we encounter the same condition
for both indices $i=1,2$, viz.,
\begin{equation}\begin{split}\label{betaLkep}
\dddot{\beta}_{L}+&\;\dot{\beta}_{L}\frac{\mu(t)}{r^{3}}+
\beta_{L}\frac{2\dot{\mu}(t)}{r^{3}}\\
\mbox{}-&\;\frac{6\mu(t)}{r^{5}}\big[b_{11}q_{1}^{2}+(b_{12}+b_{21})
\,q_{1}q_{2}+b_{22}q_{2}^{2}\big]=0\,.
\end{split}\end{equation}
We may easily identify a particular solution of this
inhomogeneous differential equation.
For $\beta_{L}(t)=0$, Eq.~(\ref{betaLkep}) is identically
satisfied for any $\vec{q}(t)$ if $b_{11}=b_{22}=0$ and
$b_{12}=-b_{21}\ne0$.
From this nontrivial solution, we get the following
contribution to the generator (\ref{lie-oper1}),
\begin{displaymath}
\boldsymbol{U}_{L,b}=q_{2}\frac{\partial}
{\partial q_{1}}-q_{1}\frac{\partial}{\partial q_{2}}\,,
\end{displaymath}
which agrees which the Noether operator (\ref{drehgen}) that
represents the conservation law of the angular momentum (\ref{dreh}).

Each fundamental solution $\beta_{L,i}(t)$, $\;i=1,2,3$
of the homogeneous part of third-order Eq.~(\ref{betaLkep}),
\begin{equation}\label{betalie}
\dddot{\beta}_{L}+\dot{\beta}_{L}\frac{\mu(t)}{r^{3}}+
\beta_{L}\frac{2\dot{\mu}(t)}{r^{3}}=0\,,
\end{equation}
is associated with the generator
\begin{displaymath}
\boldsymbol{U}_{L,\beta_{L,i}}=\beta_{L,i}(t)\frac{\partial}
{\partial t}+\onehalf\dot{\beta}_{L,i}(t)\left(q_{1}
\frac{\partial}{\partial q_{1}}+q_{2}\frac{\partial}
{\partial q_{2}}\right)\,.
\end{displaymath}
This operator formally agrees with the
corresponding operator (\ref{betagen}) of the Noether
symmetry analysis.
Yet, the respective coefficients $\beta_{i}(t)$ and
$\beta_{L,i}(t)$ are {\em different\/} in general as
they emerge as solutions from different auxiliary
equations (\ref{betakep}) and (\ref{betalie}).

For the time-independent system [$\mu(t)=\text{const}$],
the function $\beta_{L,1}(t)=1$ is a particular solution of
Eq.~(\ref{betaLkep}).
The generator $\boldsymbol{U}_{L,\beta_{L}=1}$,
\begin{displaymath}
\boldsymbol{U}_{L,\beta_{L}=1}=\frac{\partial}{\partial t}\,,
\end{displaymath}
agrees with the Noether operator (\ref{energen})
representing the energy conservation law.

A second particular solution of Eq.~(\ref{betaLkep})
for $\dot{\mu}(t)=0$ is easily shown to exist for
$\beta_{L}(t)=t$.
With this solution, Eq.~(\ref{betaLkep}) is fulfilled
identically for $b_{12}=b_{21}=0$ and $b_{11}=b_{22}=\frac{1}{6}$.
The related contribution to the generator (\ref{lie-oper1}) reads
\begin{equation}\label{gen3k}
\boldsymbol{U}_{L,\beta_{L}=t}=t\frac{\partial}{\partial t}+
\twothird q_{1}\frac{\partial}{\partial q_{1}}+
\twothird q_{2}\frac{\partial}{\partial q_{2}}\,.
\end{equation}
This generator depends on both the time and the spatial coordinates.
It thereby describes a symmetry between the spatial and the time
coordinates for this system, which reflects Kepler's third law.
We observe that the auxiliary equations of the Noether analysis
of Sec.~\ref{sec:noe-kep} do not admit a solution leading to
the generator~(\ref{gen3k}).
Therefore, the symmetry generated by Eq.~(\ref{gen3k})
is referred to as a ``non-Noether symmetry''~\cite{prince}.
Prince and Eliezer also showed that the Runge-Lenz vector
(\ref{runge-all}) can be derived on the basis of the
non-Noether symmetry generated by Eq.~(\ref{gen3k}).
This means that the variation of $I_{\psi_{i}}$ vanishes
under the action of this symmetry transformation,
\begin{displaymath}
\boldsymbol{U}^{\prime}_{L,\beta_{L}=t}I_{\psi_{i}}=0\,,
\end{displaymath}
hence that the Runge-Lenz vector constitutes an invariant
with respect to this non-Noether symmetry.
On the other hand, we have shown that the Runge-Lenz vector
in the representation of Eq.~(\ref{runge}) with $\psi_{i}(t)$
given by Eq.~(\ref{psisol}) also embodies a Noether invariant.
We conclude that the classification of the Runge-Lenz vector
as a ``non-Noether invariant'' is not justified.
This indicates once more that it is not possible to
uniquely attribute an invariant to a symmetry operator.

For $\mu(t)=\text{const}$, we can express the remaining two
independent solutions of the homogeneous part of
Eq.~(\ref{betaLkep}) in explicit form.
A comparison of Eq.~(\ref{betalie}) with the equation of
motion (\ref{kepler}) yields
\begin{displaymath}
\dot{\beta}_{L,i}(t)=c_{i}\,q_{i}(t)\,,\quad i=2,3\,,
\end{displaymath}
hence to the integral representation of $\beta_{L,i}(t)$,
\begin{equation}\label{betaL23}
\beta_{L,i}(t)=c_{i}\int_{t_{0}}^{t}q_{i}(\tau)\,d\tau\,,\quad i=2,3\,,
\end{equation}
in agreement with an approach that has been
worked out earlier by Krause~\cite{krause}.
\subsubsection{Solutions related to $\phi_{i}(t)$}
Finally, we work out the symmetries that are related to
the solutions of auxiliary equation (\ref{lie-phi}).
For the potential (\ref{keppot}) and $i=1,2$,
this equation has the particular representation
\begin{equation}\label{phikep}
\ddot{\phi}_{i}(t)-\frac{\mu(t)}{r^{5}}\Big\{3q_{i}\big[
\phi_{1}(t)\,q_{1}+\phi_{2}(t)\,q_{2}\big]-
\phi_{i}(t)\big(q_{1}^{2}+q_{2}^{2}\big)\Big\}=0\,.
\end{equation}
With $\phi_{1}(t)$ and $\phi_{2}(t)$, two linear independent
solutions of the coupled set of second-order equations (\ref{phikep}),
the related part of the generator (\ref{lie-oper1}) reads
\begin{displaymath}
\boldsymbol{U}_{L,\phi}=\phi_{1}(t)\frac{\partial}{\partial q_{1}}+
\phi_{2}(t)\frac{\partial}{\partial q_{2}}\,.
\end{displaymath}
Again, the auxiliary equations (\ref{phikep}) for the
coefficients $\phi_{i}(t)$ differ from the respective
equations (\ref{psikep}) for the coefficients
of its Noether counterpart (\ref{rlgen}).
As the subsequent solutions are different in general, we
obtain different time evolutions of the Noether and Lie symmetries.

In the case of the autonomous system with $\dot{\mu}(t)=0$,
the auxiliary equation (\ref{phikep}) possesses a simple
general solution.
Comparing Eq.~(\ref{phikep}) with the time derivative
of the equation of motion (\ref{kepler}),
\begin{displaymath}
\dddot{q}_{i}(t)-\frac{\mu}{r^{5}}\Big[3q_{i}\big(
\dot{q}_{1}\,q_{1}+\dot{q}_{2}\,q_{2}\big)-
\dot{q}_{i}\big(q_{1}^{2}+q_{2}^{2}\big)\Big]=0\,,
\end{displaymath}
we immediately see that
\begin{equation}\label{phikepsol}
\phi_{i}(t)=c\,\dot{q}_{i}(t)\,,\quad i=1,2
\end{equation}
is a solution of Eq.~(\ref{phikep}).
Herein, $c$ denotes an arbitrary constant.
For the time-independent Kepler system, the functions
of time $\phi_{1}(t)$ and $\phi_{2}(t)$ thus
coincide up to a constant factor with the time evolution
of the velocities $\dot{q}_{1}(t)$ and $\dot{q}_{2}(t)$.
We finally note the interesting result that the variation
of the Runge-Lenz vector (\ref{runge-all}) vanishes as well under
the action of the Lie symmetry generator $\boldsymbol{U}_{L,\phi}$,
\begin{displaymath}
\boldsymbol{U}^{\prime}_{L,\phi}I_{\psi_{i}}=0\,,
\end{displaymath}
which again confirms our observation that a unique correlation
between invariant and symmetry does not exist.
\subsection{Discussion}
The characteristic functions of time that are contained in both
the Noether as well as the Lie symmetry generators reflect the
specific symmetry properties of the dynamical system in question.
The Noether symmetry generator (\ref{oper1}) is determined
by the set of time functions $\beta(t)$, $\vec{\psi}(t)$,
and constants $(a_{ij})$ that follow as solutions from
differential equations (\ref{dgl1}) and (\ref{dgl2p}).
Similarly, the time functions $\vec{\alpha}(t)$, $\beta_{L}(t)$,
$\vec{\phi}(t)$, and constants $(b_{ij})$ --- following from
the auxiliary equations (\ref{alpha1}) and (\ref{lie-all}) ---
constitute the Lie symmetry generator (\ref{lie-oper1}).
The auxiliary equations for the time-dependent
coefficients of the symmetry generators generally
depend on $\vec{q}$, the system's potential $V(\vec{q},t)$,
and on the partial derivatives of this potential.
Nevertheless, particular solutions of these auxiliary equations
may exist that are decoupled from the solutions $\vec{q}(t)$
of the equations of motion.
Then, the underlying symmetry reflects a
fundamental property of the dynamical system.
As examples, we quote the solution $\beta(t)=1$ of
Eq.~(\ref{dgl1}) that exists for all cases where the
Hamiltonian (\ref{ham2}) --- hence the potential
$V(\vec{q},t)$ --- does not depend on time explicitly.
The corresponding invariant $I_{\beta}=H$
represents the energy conservation law.
Similarly, for the Kepler system the solution $a_{12}=1$
of the inhomogeneous part of Eq.~(\ref{dgl1}) provides
the fundamental law for the conservation of the angular
momentum.
Furthermore, the solution $\beta_{L}(t)=t$ of
Eq.~(\ref{betaLkep}) for $\dot{\mu}(t)=0$ reflects
the symmetry that is associated with Kepler's third law.

Nevertheless, apart from these important particular
solutions of the sets of auxiliary equations,
a wide variety of solutions exists in addition that
explicitly depends on $\vec{q}(t)$.
For instance, in the case of a Hamiltonian system
whose potential $V(\vec{q},t)$ does depend on time
explicitly, the solution $\beta(t)=1$ of
Eq.~(\ref{dgl1}) does not exist.
Apart from particular cases associated with quadratic
potentials, the solution $\beta(t)$ then depends on
the particular time evolution of $\vec{q}(t)$, and hence
on the time evolution of the potential $V\big(\vec{q}(t),t\big)$
and its partial derivatives.
As the time function $\vec{q}(t)$ is given by the
solution of the equation of motion (\ref{eqmo2}),
the solution $\beta(t)$ can only be found by integrating
Eq.~(\ref{dgl1}) {\em simultaneously\/} with the
equations of motion (\ref{eqmo2}).
This, in turn, means that the existence of the invariant
cannot help us in any way to ease the problem of
solving the equations of motion by reducing the system's order.
However, the related invariant exists and reflects
a particular symmetry of the system's time evolution.
The functional structure of the respective auxiliary
equation is well defined as it represents --- in conjunction
with the set of equations of motion --- a closed equation
with the time $t$ the only independent variable.
The dependence of a solution $\beta(t)$ on the time evolution
of $\vec{q}(t)$ does {\em not\/} indicate that the
auxiliary equations are functions of both $\vec{q}$ and $t$.
By virtue of their definition as integration ``constants,''
all coefficients of the auxiliary equations must represent
functions of time only.
We must therefore regard $\vec{q}=\vec{q}(t)$ as
time-dependent coefficients of the auxiliary equations
--- defined along the system trajectory $\vec{q}(t)$
that emerges as the solution of the equations of motion.
This requirement appears natural looking back on Noether's
theorem in the form of Eq.~(\ref{principle2}).
It indicates that the expression in brackets constitutes an
invariant if and only if the Euler-Lagrange terms of the
second sum vanish.
This is exactly the case along the system trajectory,
defined as the solution of the Euler-Lagrange equations.

In our example of the time-independent Kepler system,
we present an explicit solution (\ref{psisol}) of the
auxiliary equation (\ref{psikep}).
The functions $\psi_{i}(t)$ are understood as
functions of time satisfying Eq.~(\ref{phikep}).
Of course, we are free to identify the time evolution of the $\psi_{i}(t)$
with the {\em time evolution\/} of the particle coordinates
$q_{i}(t)$ and $\dot{q}_{i}(t)$, or combinations thereof.
With this understanding, the functions $\psi_{i}(t)$
remain functions of time only --- which is
crucial for Eq.~(\ref{u-psi}) to be satisfied.
The Runge-Lenz vector of the Kepler system can then be
conceived as a Noether invariant, generated by a particular
operator of the type $\boldsymbol{U}_{\!\psi}$.
\section{Conclusions}
We have worked out in detail the Noether and Lie symmetry
analyses for a Hamiltonian comprising the explicitly
time-dependent general potential $V(\vec{q},t)$.
In this general form, the analyses resulted in specific
sets of ordinary differential equations for the coefficients
of the symmetry generators.
The search for Noether and Lie symmetries could thus be
reduced to the pursuit of the complete variety of solutions
of the sets of auxiliary equations.

The auxiliary equations of the Lie approach were found
to agree with the partial $q_{i}$ derivative of the
corresponding Noether equations.
For one, this indicates the close relationship between
these two approaches.
On the other hand, the sets of auxiliary equations
are obviously different, hence, apart from particular
isotropic systems associated with quadratic potentials,
the solution functions are different in general.
This means that the time evolutions of Noether and Lie
symmetries generally do not agree, hence that the set of
Noether symmetries cannot be regarded as a subset of the
Lie symmetries.

For the Noether approach, we have seen that there exists a
one-to-one correspondence between symmetry and a related invariant.
The reason for this is that the Noether auxiliary equations
--- in conjunction with the equations of motion ---
can always be cast into the form of a total time derivative.
This does not hold for the auxiliary equations that emerge
from the Lie symmetry analysis.
Therefore, the Lie symmetries are not necessarily associated
with closed-form expressions of conserved quantities.

Depending on their specific form for a given potential,
the Noether auxiliary equations (\ref{dgl1}) and
(\ref{dgl2p}) as well as the corresponding Lie auxiliary
equations (\ref{alpha1}) and (\ref{lie-all}) may have
explicit solutions, which then reflect fundamental
symmetries of the dynamical system in question.
Particular solutions of these equations may exist as well
that decouple from the system trajectory $\vec{q}(t)$
representing the solution of the equations of motion as
the system moves forward in time.
On the other hand, additional solutions of the auxiliary
equations exist that explicitly depend on the evolution
of $\vec{q}(t)$.
These solutions must be taken into consideration as well
in order to obtain the full set system symmetries.

With this perception of the auxiliary equations, all invariants
of the Kepler system could be derived from Noether's theorem.
In particular, the Runge-Lenz vector has been identified
as a Noether invariant.
In this regard, we confirm the claim made by Sarlet and
Cantrijn~\cite{sarlet} that ``all integrals of Lagrangian
systems can indeed be found by a systematic exploration
via Noether's theorem.''
This contrasts with the statement of Prince and Eliezer~\cite{prince}
that ``the Runge-Lenz vector evades detection by Noether's theorem.''
The reason for this discrepancy originates in a different
understanding of the coefficients of the auxiliary equations.
By virtue of their definition being functions of time only, the
coefficients must not depend on $\vec{q}$ and $\dot{\vec{q}}$.
Nevertheless, we are free to relate the {\em time evolution\/}
of these coefficients with the time evolution of $\vec{q}(t)$
and $\dot{\vec{q}}(t)$.
With this enhanced understanding of the solutions of the
auxiliary equations, a broader solution ``spectrum'' is
obtained, which results in a wider range of invariants
emerging within the framework of Noether's theorem.
As a consequence, yet unknown nonlocal Noether symmetries
of the Kepler system could be isolated.

Furthermore, we have worked out additional solutions of the
Lie auxiliary equations for the Kepler system that yield
yet unreported symmetries of this system.

We have seen that the variations of the obtained invariants vanish
with respect to different Noether as well as Lie symmetry operators.
This demonstrates that a unique correspondence between an
invariant and a symmetry transformation does not exist.
As a consequence, a classification of invariants with respect
to a certain symmetry operation is not possible.

Analyzing the symmetries of a given dynamical system,
the conditions under which certain solutions of the auxiliary
equations cease to exist can be identified as the regions in
parameter space where the related symmetries disappear.
This way, we may efficiently isolate the causes that render a
dynamical system less symmetric, hence more chaotic or even unstable.


\begin{thebibliography}{}
\bibitem{lie} S.~Lie, {\it Vorlesungen \"uber Differentialgleichungen}
  (Teubner Verlag, Leipzig, 1891); see also, for instance, P.~J.~Olver,
  {\it Applications of Lie Groups to Differential Equations}
  (Springer Verlag, New York, 1986).
\bibitem{noether18}
  E.~Noether, Nachr.~Ges.~Wiss.~G\"ottingen,
  Math.-Phys.~Kl.~{\bf 57}, 235 (1918).
\bibitem{struck-rie00}
J.~Struckmeier and C.~Riedel, Phys.~Rev.~Lett.~{\bf 85}, 3830 (2000).
\bibitem{struck-rie01}
J.~Struckmeier and C.~Riedel, Phys.~Rev.~E~{\bf 64}, 026503 (2001).
\bibitem{struck-rie02}
J.~Struckmeier and C.~Riedel, Ann.~Phys.~(Leipzig)~{\bf 11}, 15 (2002).
\bibitem{stephani}
  H.~Stephani, {\it Differential Equations},
  (Cambridge University Press, Cambridge, 1989).
\bibitem{hill51}
  E.~L.~Hill, Rev.~Mod.~Phys.~{\bf 23}, 253 (1951).
\bibitem{lutzky78a}
  M.~Lutzky, Phys.~Lett.~{\bf 68A}, 3 (1978).
\bibitem{ermakov}
  V.~P.~Ermakov, Universitetskiye~Izvestiya~Kiev~{\bf 20}, 1 (1880)
  (in Russian).
\bibitem{leach-lewis82}
  H.~R.~Lewis and P.~G.~L.~Leach, J.~Math.~Phys.~{\bf 23}, 2371 (1982).
\bibitem{wulf}
  C.~E.~Wulfman and B.~G.~Wybourne, J.~Phys.~A~{\bf 9}, 507 (1976).
\bibitem{prince}
  G.~E.~Prince and C.~J~Eliezer, J.~Phys.~A~{\bf 14}, 587 (1981).
\bibitem{krause}
  J.~Krause, J.~Math.~Phys.~{\bf 35}, 5734 (1994).
\bibitem{sarlet}
  W.~Sarlet and F.~Cantrijn, SIAM~Rev.~{\bf 23}, 467 (1981).
\end{thebibliography}
\end{document}